\pdfoutput=1
\documentclass[11pt,twoside]{article}  
\usepackage{epsfig}

\usepackage{geometry}
\newcommand{\beq}{\begin{eqnarray*}}
\newcommand{\eq}{\end{eqnarray*}}

\newcommand{\var}{\mbox{Var }}

\newcommand{\bdis}{\begin{displaymath}}
\newcommand{\edis}{\end{displaymath}}

\newcommand{\rP}{\mbox{P}}

\begin{document}

\title{A powerful allele based test for case-control association studies}
\author{M.A. Jonker$^1$, M.W.T. Tanck$^2$\\ \\
$^1$VU Medical Center, Amsterdam, The Netherlands.\\
$^2$AMC Amsterdam, The Netherlands\\
m.jonker2@vumc.nl, m.w.tanck@amc.uva.nl}
\date{}
\maketitle

\begin{abstract}
{In a case-control study aimed at localizing disease variants, association between a marker and the 
disease status is often tested by comparing the marker allele frequencies among cases and controls. 
These marker allele frequencies are expected to be different if the marker is associated 
with the disease. The power of the commonly used allele based test is based on 
the marker allele frequency; markers with a low minor allele frequency have less power to be detected 
(if they are associated with the disease), than markers with high minor allele frequency. 
Therefore the strategy of selecting markers for follow-up study based on their p-values, favors markers with a
high minor allele frequency.  

We propose an 
allele based test that does not have this (unwanted) property and is therefore more powerful for markers 
with a low minor allele frequency. This test may, therefore, be more 
effective when searching for rare causal variants. The asymptotic power function of the test is derived and
simulation studies are performed for finite sample properties of the test. 
Next, the existing and the proposed tests are applied to data; this is not included yet. 

In the light of the current interest in detecting association between complex phenotypes and causal variants with 
a low minor allele frequencies, this test is expected to be of relevance.}
\end{abstract}

\medskip

\noindent
Case-control study, allele based test, linkage disequilibrium (LD), power, p-values, minor allele frequency (MAF)

\newpage

\section{Introduction}
To locate disease variants and dichotomous trait loci, association studies of genetic 
markers are often conducted with a case-control design. Several genetic association tests
have been proposed (see e.g. \cite{Balding2006,BookAsso}). 
One simple and perhaps one of the most natural test, is the singular marker allele based test 
that is based on the difference of the sample marker allele frequencies
in cases and controls. 

A marker is associated with the disease or trait if it is in linkage disequilibrium (LD) 
with one of the causal variants (see e.g. \cite{BookAsso}).
On average, the larger the degree of LD, the smaller the p-value of the association test.
However, the p-value obtained with the (commonly used) singular marker allele based test is 
also a function of the marker allele frequency; 
markers with high minor allele frequencies (MAFs) have more power to be detected than markers with low MAFs. 
The strategy of selecting interesting markers for follow-up study based on their p-values only, is therefore biased towards 
markers with high MAFs. 

Several other strategies for prioritizing the markers for follow-up studies have been proposed and compared,
like ranking markers based on the Bayes Factor signal (\cite{Wakefield2008,Wakefield2009}), the likelihood ratio signal (\cite{Stromberg2009}), 
frequentist factor signal (\cite{W2004}), and PrPES signals (\cite{Stromberg2008}).
\cite{Stromberg2009} compared these strategies, including ranking markers based on the p-values of the allele based test and 
Cochran-Armitage-trend test, by applying them to two data-sets.
The markers with the smallest p-values obtained from the allele based test are also top-ranked by the other methods. 
Some strategies down-weight markers with small MAF even more than the allele based test does. 

In this paper we propose an allele based test for testing association in case-control studies that
does not favor markers with high minor allele frequencies. It is shown that the test has more power than the 
commonly used allele based test if the minor allele frequency of the marker is quite low. 
The proposed test-statistic is found by standardizing the difference of sample allele frequencies in a different way
than the commonly allele based test does. An explicit asymptotic power function is derived 
and finite sample properties are obtained by simulation studies.
The test will be applied to data. 

``Missing'' heritability refers to the fact that for many traits, only a small
proportion of the variability in the population can be explained by causal variants 
that have been identified up to now (see e.g. \cite{Manolio}).   
One possible explanation for this ``missing'' heritability is 
the presence of low-frequency variants of relatively strong effect on disease risk. 
Indeed, rare variants found by resequencing have already been 
described to affect complex diseases (\cite{Schork}). In the light of the current
interest in detecting association between complex phenotypes and low-frequency 
variants and localizing causal variants with small minor allele frequencies,
the implications of the present paper are expected to be of relevance.

\section{Methods}
\subsection{Setting}
The case-control status for a random individual in the general population is denoted by $X$,
$X=1$ for a case and $X=0$ for a control. The fractions of cases and controls in the total population 
are denoted by $\pi=\rP(X=1)$ and $1-\pi=\rP(X=0)$. There may be multiple causal variants; the aim in 
association studies is to locate these variants.
We initially assume that any given marker is in linkage disequilibrium with at most 
one causal variant. Therefore, it is sufficient to consider the situation where there is only 
one causal variant. 
Suppose the causal variant is biallelic with alleles $A_1$ and $A_2$ and with corresponding allele frequencies 
$p_1$ and $p_2=1-p_1$ in the total population. In case the causal variant is not biallelic, the second allele, $A_2$, 
could be regarded as all alleles which are not the $A_1$-allele. 
We denote the fraction of individuals with the disease ($X=1$) among those individuals with genotype $(A_i,A_j)$ at the 
causal variant by $\pi_{ij}=\rP(X=1|A_iA_j)$ for $i,j=1,2$ and 
we assume that $\pi_{12}=\pi_{21}$.

We consider a biallelic marker which may or may not be in proximity with the causal variant. 
One of the marker alleles is denoted as $M_1$ and the other allele as $M_2$.
The allele frequencies for $M_1$ and $M_2$ in the general population are denoted
as $q_1$ and $q_2=1-q_1$ and the $M_i$ allele frequencies among the controls and cases 
are denoted as $q_{i|0}=q_{i|X=0}$ and $q_{i|1}=q_{i|X=1}$ for $i=1,2$. 

A common measure for the degree of LD between a marker and a causal variant 
is given by the quantity $\Delta_{ij}=D_{ij}/\sqrt{p_1p_2q_1q_2}$, with $D_{ij} = \rP(A_iM_j) - p_i q_j$ where 
$\rP(A_iM_j)$ is the $(A_i,M_j)$ haplotype-frequency in the total population (see for instance \cite{Devlin}). 
By definition $\Delta_{ij}=\mbox{cor}(1_{A_i},1_{M_j})$ 
with, for a randomly chosen haplotype, $1_{A_i}$ and $1_{M_j}$ the indicator functions which equal 1 if 
the causal variant is $A_i$ and 0 otherwise, and similar for the marker allele $M_j$.
In the Appendix A in the Supplementary Material it is derived that
\begin{eqnarray}
\frac{q_{1|0} - q_{1|1}}{\sqrt{q_1q_2}} = \Delta \; \frac{p_{1|0} - p_{1|1}}{\sqrt{p_1p_2}}
\label{pMLUDelta}
\end{eqnarray}
with $\Delta=\Delta_{11}$.
So, the relative difference between the allele frequencies among the controls and cases at the 
causal variant (the quotient on the right hand side of the expression) is passed on to the neighboring markers by multiplying
this relative difference by $\Delta$, the degree of linkage disequilibrium between the alleles at the marker and the causal variant. 
From this formula it can be directly seen that the $M_1$ allele frequencies among the controls and cases equal
if the marker is in linkage equilibrium with the causal variant, i.e. $\Delta=0$.

In order to find markers that are associated with the disease or, actually, are in linkage disequilibrium with the causal variant,
case-control data is collected. Suppose we have a sample of $R$ individuals from the cases and $S$ individuals from the controls; 
$R$ and $S$ are fixed and non-random. Their sum is denoted as $N=S+R$.
Since every genotype consists of two alleles, there are actually $2R$ and 
$2S$ alleles from the cases and controls, respectively.
The number of $M_1$ alleles among the cases and controls are denoted as $R_1$ and $S_1$, respectively.  
For the number of $M_2$ alleles, the notation is analogues: $R_2$ and $S_2$ for the cases and controls.
Note that $R_1+R_2=2R$ and $S_1+S_2=2S$. 
Based on these data the fraction of $M_1$ alleles among the controls and the cases can be estimated as
$\hat{q}_{1|0} = S_1/(2S)$ and $\hat{q}_{1|1} = R_1/(2R)$.

In the following two allele based association tests are described
for testing the null hypothesis H$_0: q_{1|0} = q_{1|1}$ against the alternative H$_1: q_{1|0} \neq q_{1|1}$.
The first test we describe, is the commonly used test. This test is based on the difference 
$\hat{q}_{1|0} - \hat{q}_{1|1}$,
standardized so that the test-statistic is asymptotically
standard normal distributed under the null hypothesis of no association. 
Calculations will show that by the way of standardizing, markers with a low
minor allele frequency have less power to be detected under the alternative hypothesis, than markers with a 
high minor allele frequency. The test we propose is also based on
the difference $\hat{q}_{1|0} - \hat{q}_{1|1}$, but standardized in a different way. 
In Section 3 the power of the two tests are compared.

\subsection{Commonly used allele based association test}
The commonly used allele based test-statistic for testing the null hypothesis H$_0: q_{1|0} = q_{1|1}$ against
H$_1: q_{1|0} \neq q_{1|1}$ is given by
\beq
T = \frac{\hat{q}_{1|0} - \hat{q}_{1|1}}{\sqrt{\hat{V}}}
\eq
with $\hat{q}_{1|0} = S_1/(2S)$ and $\hat{q}_{1|1} = R_1/(2R)$.
Furthermore, $\hat V$ is an estimator of the variance $V = \var (\hat{q}_{1|0} - \hat{q}_{1|1}) = V_{0} + V_{1}$,
with $V_0$ and $V_1$ defined as 
\begin{eqnarray}
V_{0} = \var \hat{q}_{1|0} = \frac{q_{1|0}(1-q_{1|0})}{2S} \qquad
V_{1} = \var \hat{q}_{1|1} = \frac{q_{1|1}(1-q_{1|1})}{2R}.
\label{VXL}
\end{eqnarray}
The unknown frequencies $q_{1|0}$ and $q_{1|1}$ in $V_0$ and $V_1$ are, usually, estimated 
by $\hat{q}_{1|0}$ and $\hat{q}_{1|1}$, so that $V$ is asymptotically unbiased under the null and alternative hypothesis. 
For $\lambda = R/N$ the fraction of sampled cases, the variance $V$ equals
\beq
V = V_0 + V_1 = \frac{\lambda q_{1|0}q_{2|0} + (1-\lambda)q_{1|1}q_{2|1}}{2N\lambda(1-\lambda)}  = m^{-1} (\lambda q_{1|0}q_{2|0} + (1-\lambda)q_{1|1}q_{2|1})
\eq
for $m=2N\lambda(1-\lambda)$.
For large sample sizes, the test-statistic $T$ has, approximately, a normal distribution with mean  
\begin{eqnarray}
\sqrt{m}\; \frac{q_{1|0} - q_{1|1}}{\sqrt{\lambda q_{1|0}q_{2|0}+(1-\lambda) q_{1|1}q_{2|1}}} \; = \; \sqrt{m} Q B \Delta
\label{meanT}
\end{eqnarray}
and variance 1, where
\begin{eqnarray}
B = \frac{p_{1|0} - p_{1|1}}{\sqrt{p_1p_2}} \qquad \mbox{ and } \qquad Q^2 = \frac{q_1q_2}{\lambda q_{1|0}q_{2|0}+(1-\lambda) q_{1|1}q_{2|1}}.
\label{BQ}
\end{eqnarray}
The equality in (\ref{meanT}) follows from (\ref{pMLUDelta}).
Under the null hypothesis that $q_{1|0} = q_{1|1}$ (i.e. $\Delta=0$),   
$T$ has, asymptotically, a standard normal distribution, whence $H_0$ is rejected for $|T| \geq z_{\alpha/2}$,
with $z_{\alpha/2}$ the upper-$\alpha/2$-quantile of the standard normal distribution.
The two-sided power-function of the test (approximately) equals, for large sample sizes,
\begin{eqnarray}
\rP(|T| \geq z_{\alpha/2}) =1-\Phi(z_{\alpha/2}-\sqrt{m} B \Delta Q) + \Phi(-z_{\alpha/2}-\sqrt{m} B \Delta Q).
\label{powT}
\end{eqnarray}
The power of the test is controlled by the product $\sqrt{m} B \Delta Q$.
The first term, $\sqrt{m}$, is specific for the way of sampling and is, therefore, equal for every marker.
The second term, $B$, is specific for the causal variant and is equal for all markers
that are in linkage disequilibrium with the same causal variant. 
The third term, $\Delta$, measures the degree of LD between the marker and the causal variant and its value varies 
over the markers.
The last term, $Q$, depends on the marker. Under the null hypothesis that a marker is not associated with the disease $Q=1$, 
but under the alternative $Q$ is either smaller or larger than 1, depending on the (conditional) marker allele frequencies. 
That means that the markers in the association study are weighted; markers with $Q>1$ do get more power to be detected than
markers with $Q<1$. 


\subsection{Allele based test for markers with low MAF }
Define the test-statistic  
\begin{eqnarray}
W = \sqrt{m} \; \frac{\hat q_{1|0} - \hat q_{1|1}}{\sqrt{\hat q_1 \hat q_2}}
\label{W}
\end{eqnarray}
with $m=2N\lambda(1-\lambda)$ (as before) and $\hat q_1 = \hat \pi \hat{q}_{1|1} + (1-\hat\pi) \hat{q}_{1|0}$
and $\hat q_2 = \hat \pi \hat{q}_{2|1} + (1-\hat\pi) \hat{q}_{2|0}$, where $\hat\pi$
is an estimate of the disease prevalence $\rP(X=1)$. This estimate cannot be obtained from 
the samples of cases and controls, but should be estimated based on external data. For many diseases, 
an estimate of the population prevalence $\pi$ is available, for instance in the literature or in national registries.

The test-statistics $W$ and $T$ are related via the relationship $W = T \hat Q^{-1} $, with $\hat Q$ an estimate 
of $Q$ in (\ref{BQ}); it is found by inserting the estimates $\hat q_1$ and $\hat q_2$ (as just defined) 
and the marker sample frequencies among cases and controls.
By the law of large numbers, $\hat Q$ approximately equals $Q$ for large samples and
from Slutsky's lemma and the continuous mapping theorem (see e.g. \cite{vanderVaart}) it follows that $W$ has, 
approximately, a normal distribution with mean 
\beq
\sqrt{m}\; \frac{q_{1|0} - q_{1|1}}{\sqrt{q_1 q_2}} = \sqrt{m} B\Delta,
\eq
and variance $Q^{-2}$.
Under the null hypothesis, $q_{1|0}=q_{1|1}, \Delta=0$ and $Q=1$ and $W$ has, asymptotically, 
a standard normal distribution, whence $H_0$ is rejected for $|W| \geq z_{\alpha/2}$ for $z_{\alpha/2}$
the upper $\alpha$ quantile of the standard normal distribution.
Under the null hypothesis 
$\hat q_1 = \hat \pi {q}_{1|1} + (1-\hat\pi) {q}_{1|0} \approx \hat \pi {q}_{1} + (1-\hat\pi) {q}_{1} =q_1$,
no matter the value of $\hat\pi$. That means that, even if the estimate of $\pi$ is far away from the true value, the 
type I error of the test will still be approximately correct if the sample sizes are large enough. 
The two-sided power function for $W$ is given by
\begin{eqnarray}
\rP(|W| > z_{\alpha/2}) &\approx & \rP(|T| > z_{\alpha/2}Q) \nonumber\\
&=& 1-\Phi(z_{\alpha/2}Q-\sqrt{m} B \Delta Q) + \Phi(-z_{\alpha/2}Q-\sqrt{m} B \Delta Q). 
\label{powW}
\end{eqnarray}
The power functions for $W$ and $T$ are very similar, but differ in the way $Q$ is the expression.
In the power function for $W$, also the quantile $z_{\alpha/2}$ is multiplied with $Q$. If $Q=1$ (i.e. under $H_0$)
the power functions equal (to $\alpha$).
Under the alternative hypothesis either $Q>1$ or $Q<1$. If the sample size is large, this will probably also hold 
for $\hat Q$ (since $\hat Q$ converges in probability to $Q$ if also $\hat\pi$ converges in probability to $\pi$). 
From the definitions of $T$ and $W$, it follows that $|W| > |T|$ if $\hat Q<1$ and $|W| < |T|$ if $\hat Q>1$; 
if $\hat Q<1$ the test based on $W$ is more powerful, whereas the test based on $T$ is more powerful if $\hat Q > 1$.

In Appendix B in the Supplementary Materials it is 
shown that, if $M_1$ is positively correlated with the risk allele ($A_1$ or $A_2$)
and $q_1$ is sufficiently small, $Q$ will be smaller than 1. However, if $M_1$ is negatively correlated with the
risk allele, $Q>1$.   
If the risk allele is the minor allele and $q_1$ is small, strong negative correlation with the risk allele are not 
possible within the parameter space
of the genetic model. The latter can be easily seen from the following.
Remind that $\Delta = (\rP(A_1,M_1)-p_1q_1)/\sqrt{p_1p_2q_1q_2}$.
Consequently, $\Delta \geq -p_1q_1/\sqrt{p_1p_2q_1q_2}$.
If $p_1=q_1=0.05$, $\Delta> -0.053$, and $p_1=0.25, q_1=0.05$ yields $\Delta > -0.12$
and $p_1=q_1=0.25$, yields $\Delta>-0.26$.

\subsection{Generalization of the test-statistic: $W_\delta$}
The test-statistic $W$ can be generalized by allowing other values in stead of $\pi$. Define 
the test-statistic 
\begin{eqnarray}
W_\delta =  \frac{\sqrt{m}(\hat q_{1|0} - \hat q_{1|1})}{\sqrt{(\delta \hat{q}_{1|1} + (1-\delta) \hat{q}_{1|0})(\delta \hat{q}_{2|1} + (1-\delta) \hat{q}_{2|0})}}.
\label{Wdelta}
\end{eqnarray}
For $\delta=\hat\pi$, this statistic $W_{\hat\pi}$ equals $W$ and, furthermore,
$W_\delta = T \hat Q^{-1}_{\delta}$ with $\hat Q_{\delta}$ an estimate for $Q_\delta$ with 
$Q^2_\delta = q_1q_2/(\delta q_{1|0}q_{2|0}+(1-\delta) q_{1|1}q_{2|1})$ (similar to $Q^2$ as defined before).
The estimate of $Q_\delta$ is found by inserting estimates of the allele frequencies, like is done for $Q$.

Under the null hypothesis that $q_{1|0}=q_{1|1}$ and the denominator of $W_\delta$ 
converges in probability to $\sqrt{q_1q_2}$, notwithstanding the value of $\delta$.
That means that under the null hypothesis, the test-statistic $W_\delta$ has, approximately, a standard normal distribution,
for large sample sizes. So, the null hypothesis of no association is rejected if $|W_\delta| > z_{\alpha/2}$.

If $\delta=\pi$, the denominator converges to $\sqrt{q_1q_2}$ under the alternative hypothesis.
So, only in that case the mean of the test-statistic approximates $\sqrt{m} B \Delta$ for large sample sizes.
The test is most powerful if, under the alternative hypothesis, the denominator is minimized.
If $M_1$ is the minor allele and positively correlated with the disease and, thus $q_{1|0}<q_{1|1}<0.5$, the test is optimal
for $\delta=0$ and least optimal for $\delta=1$. 
If $0.5 > q_{1|0}>q_{1|1}$ (the minor allele $M_1$ is negatively correlated with the disease), it is the other way around.
For $0<\delta<1$, the power is in between the minimum and maximum.
In practice it is unknown whether the minor allele $M_1$ is positively or negatively correlated with the 
disease.

\subsection{Combined test of $W$ and $T$}
The two test-statistics $W$ and $T$ are linked via $T = W \hat Q$. If $\hat Q>1$, the test with test-statistic $T$ is more powerful, 
whereas the opposite holds if $\hat Q<1$. If $\hat Q=1$ then $T=W$ and the two tests are equivalent.
To have best of both of tests, the two tests could be combined. Define 
\begin{eqnarray}
U = T 1_{\hat Q > 1} + W 1_{\hat Q \leq 1} = W \hat Q 1_{\hat Q > 1} + W 1_{\hat Q \leq 1}   
\label{U}
\end{eqnarray}
as the combined test-statistic. 
By combining the law of large numbers, Slutsky's lemma and the continuous mapping theorem (see e.g. \cite{vanderVaart}),
$U$ has asymptotically a standard normal distribution under the null hypothesis of no association;
the null hypothesis is rejected for $|U| \geq z_{\alpha/2}$.
The asymptotic power function for $U$ equals the one of $W$ if $Q<1$ and of $T$ if $Q>1$.

In a similar way the test-statistics $W_\delta$ and $T$ could be combined.

\subsection{Multiple causal variants}
Suppose there are multiple causal variants, but every marker is in linkage disequilibrium with at most one causal variant.
The value of $B$ is specific for the causal variant and will, therefore, be the same for all 
markers which are in linkage disequilibrium with this causal variant.
Since the power function (and the p-value) depends on the value of $B$, markers can only be ranked on their p-values locally; 
for all markers which are in linkage disequilibrium with the 
same causal variant. In practice it is unknown whether there is only one or multiple causal variants. One should be careful
when comparing p-values for markers at different chromosomes or located far apart.

\subsection{Measuring the effect size}
For the markers that show significant association with the disease status, the effect size is of interest. 
When testing based on the test-statistics $T$ or $W$, the difference between the  
allele frequencies among cases and controls, $q_{1|0}-q_{1|1}$, could be used as a measure of effect, but 
this difference is difficult to interpret. In practice the Cochran-Armitage test or the test-statistic $T$, 
is often used for testing association, whereas an effect size in terms of odds ratios is estimated by fitting 
a logistic regression model (\cite{Wellek2012}). 
It would be more natural to use the same statistic for testing association 
and estimating an effect size. This can be done within a logistic regression model, but also 
based on the test-statistics $W$ or $T$. The effect size defined as
\beq
\frac{\rP(X=0|M_1)}{\rP(X=1|M_1)} \Bigg/ \frac{\rP(X=0)}{\rP(X=1)}, 
\eq
can be estimated and a corresponding confidence interval can be constructed. 
This can be seen by writing this fraction as (using Bayes theorem) $q_{1|0}/q_{1|1} = 1+(q_{1|0}-q_{1|1})/q_{1|1}$ and
noting that $\hat{q}_{1|1}$ is consistent by the law of large numbers. 
This quantity is not based on any model assumptions, like in a logistic regression model, 
and is therefore very appropriate for quantifying the marker effect size. 

When determining a quantity for measuring an effect size, it is important to keep in mind what the 
aim of the study is. In case one aims to estimate the disease risk based on the observed genotypes at markers which
are possibly not the causal variant, odds ratios for different markers would be appropriate. However, in case one aims to find
the causal variant, one may prefer to use a measure that tries to quantify
the distance between markers and a causal variant, or at least orders the markers with respect to their distance to
the causal variant. Of course, it is not possible to measure the physical distance between markers and a causal variant,
but, locally, the markers can be ordered with respect to the degree of linkage disequilibrium with a causal variant.      
In the previous subsection we have seen that $W/\sqrt{m} \approx B \Delta$, where $B$ is equal for all markers 
that are in LD with the same causal variant. That means that 
ranking the markers (in a small region of the chromosome) based on the test-statistic $|W|$, 
is equivalent to ranking them based on (an estimate of) the degree the marker is in LD with the causal variant ($\Delta$). 
Although, it is known that $\Delta$ between markers and a causal variant is not necessarily a monotone function 
with the physical distance between the two, it is expected that there is a positive relationship
and the causal variant will be located nearby the markers which are strongest in LD with the causal variant.

\section{Results}
This section is divided into two subsections. In the first subsection the asymptotic power functions of the two tests are 
compared. In practice, the sample sizes are finite. In the second subsection we perform simulation to study the type I error
for finite samples, and 
the effect of the extra variability due to the fact that $\pi$ is estimated.

\subsection{Comparison of the asymptotic power functions}
We define the $A_1$-allele at the causal variant as the high risk allele 
and the disease probabilities are taken equal to $\pi_{11} = 0.60, \pi_{22} = 0.10$ and 
for the additive model $\pi_{12} = (\pi_{11}+\pi_{22})/2=0.35$.

In Figure 1 the asymptotic power function is given as a function of $q_1$ (first and 
second plot) and of $\Delta$ (third plot). In all cases $R=S=1000$ and $\alpha = 1.0e^{-8}$. 
In the first plot $p_1=0.05$ and $p_1=0.15$ in the second plot. In both cases $\Delta=0.3$.  
The power function for the test-statistic $W$ is plotted as a continuous line and 
for $T$ as a dashed line. 
In the most right plot in Figure 1, the power function as a function of $\Delta$ is given.
Now, $p_1=0.05$ and $q_1$ equals either $p_1$ or $3p_1$.
Again, the dashed lines represent the power for the test with test-statistic $T$ ($q_1=p_1$ lower line,
$q_1=3p_1$ upper line) and the continuous lines for the test with test-statistic $W$ (the two lines overlap).
For all plots we inserted the true value of $\pi$ in the power function, because it will be shown below 
that the power function is robust against misspecification of the parameter $\pi$.

From the plots it can be concluded that the test based on test-statistic
$W$ is more powerful than the one based on $T$ for the genetic models described above.
The power based on $W$ is approximately constant as a function of $q_1$, whereas
the power based on $T$ increases with $q_1$.
So, $W$ does not, a-priori, favors markers with a large minor allele frequency.
This makes the p-values comparable across markers.
This does not hold for $T$. 
The power functions were plotted for more genetic models, including the dominant and the recessive model.  
As long as the correlation between $A_1$ and $M_1$ is positive, the conclusions remain the same. 
\begin{figure}[!p]
\centering
\includegraphics[height=2.5in,width=6.5in]{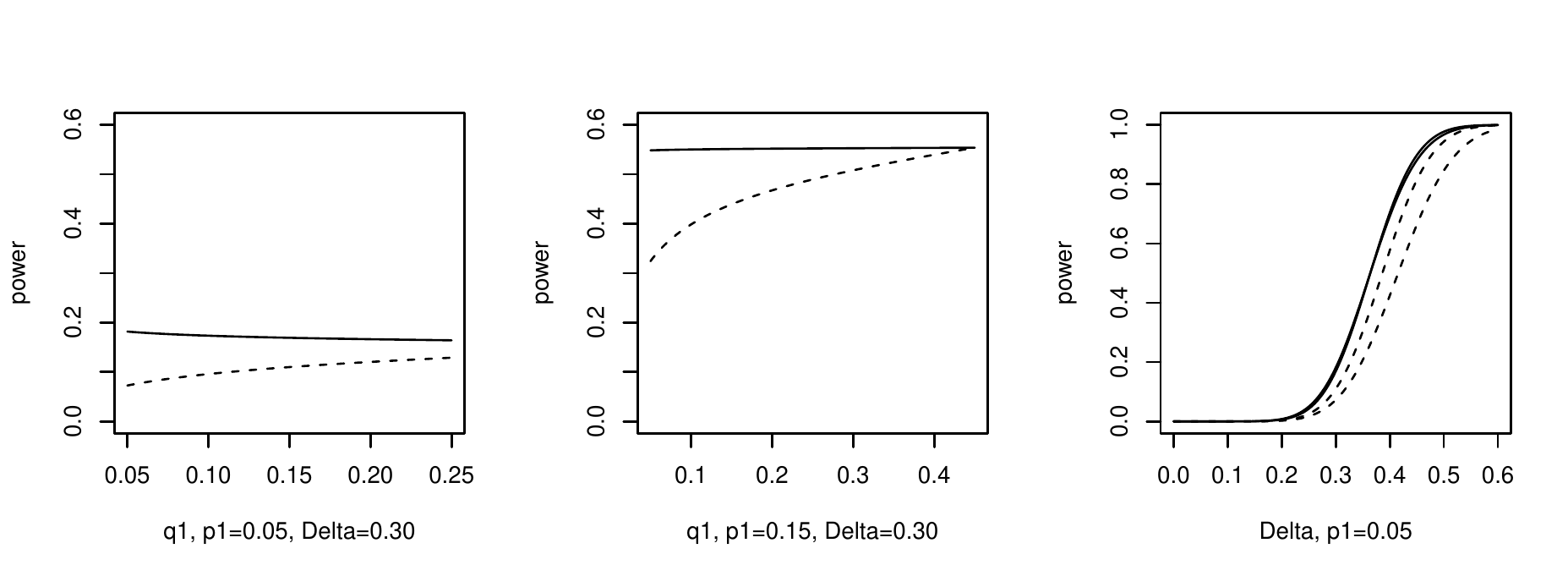}
\caption{Power functions for $T$ (dashed lines) and $W$ (continuous lines). 
Additive model with: $\pi_{11} = 0.60, \pi_{22} = 0.10$ and $\pi_{12} = 0.35$.}
\end{figure}

If the minor allele $M_1$ is negatively correlated with the causal allele $A_1$, the theory tells us that
the test based on $T$ is more powerful. We consider the same setting as before.  
In the left plot of Figure 2 the power functions for the two tests are given as a function of $q_1$ 
for $\Delta= -0.40$ and $p_1=0.60$. The power of the test based on $T$ is indeed higher.
However, note the low power of both tests. In the right plot of Figure 2 the power is plotted as a function of $\Delta$.
Again, the test $T$ is more powerful if $\Delta<0$.   
\begin{figure}[!p]
\centering
\includegraphics[height=2.5in,width=2.5in]{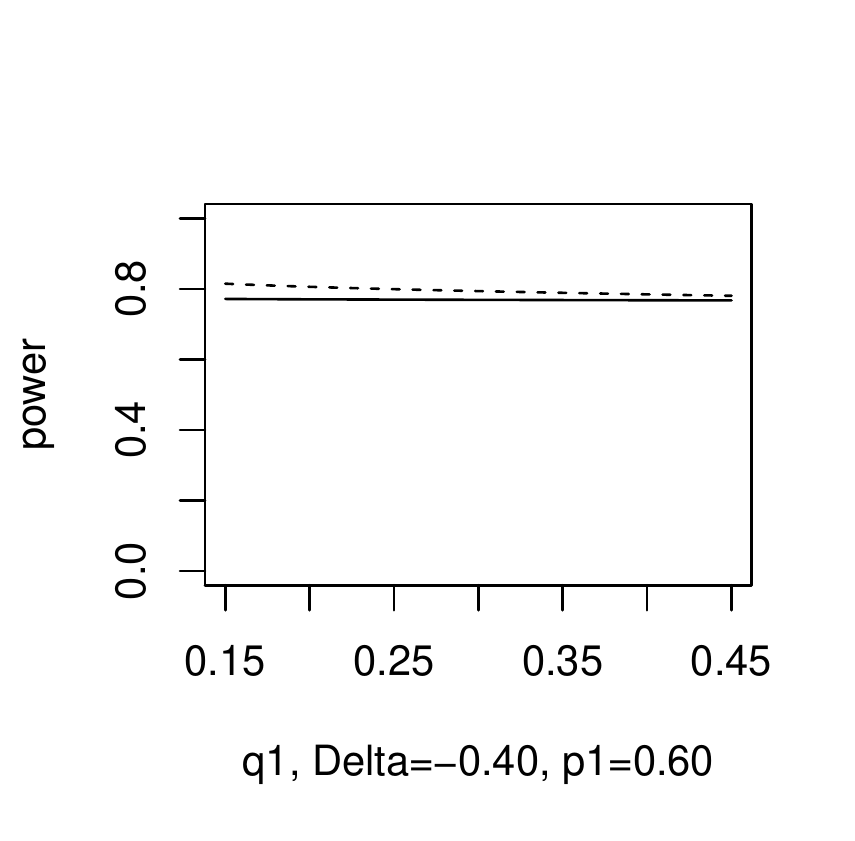}
\caption{Power functions for $T$ (dashed lines) and $W$ (continuous lines) in the additive
model with, left $\Delta=-0.40$.}
\end{figure}

\bigskip

\noindent
{\underline{Misclassification of $\pi$}}\\
The denominator of the test-statistic $W$ contains the parameter $\pi$, the prevalence of the disease,
which cannot be estimated from the case-control data
itself, but has to be estimated based on data from a different source. 
Misspecification of this parameter affects the power of the test. Therefore, the power function is considered 
for different values of the estimate $\hat\pi$. So, actually the power of the test based on $W_\delta$ is considered for
different values for $\delta$ near $\pi$. That means that misclassification of $\pi$ will never lead to a test with 
inflated type I error (if the sample size is big enough), but to a different test which has a priori a slight 
preference for markers with either small (if $\hat \pi < \pi$) or large (if $\hat \pi > \pi$) minor allele frequencies.

The power function was computed for exactly the same models as
was done before. The results are given in Figure 3.
The power was computed for $\hat\pi = 0.2$ (lowest continuous lines)
$\hat \pi = \pi = 0.125$ (continuous line in the middle), $\hat\pi = 0.075$ (upper line). 
Since the power function for $T$ does not contain $\pi$, only one curve is found; the dashed line.   
More models were considered. In all cases similar results were obtained. 
We conclude that allele based test for testing with the test-statistic $W$ is robust against small misspecification
of $\pi$.

\begin{figure}[!p]
\centering
\includegraphics[height=2.5in,width=2.5in]{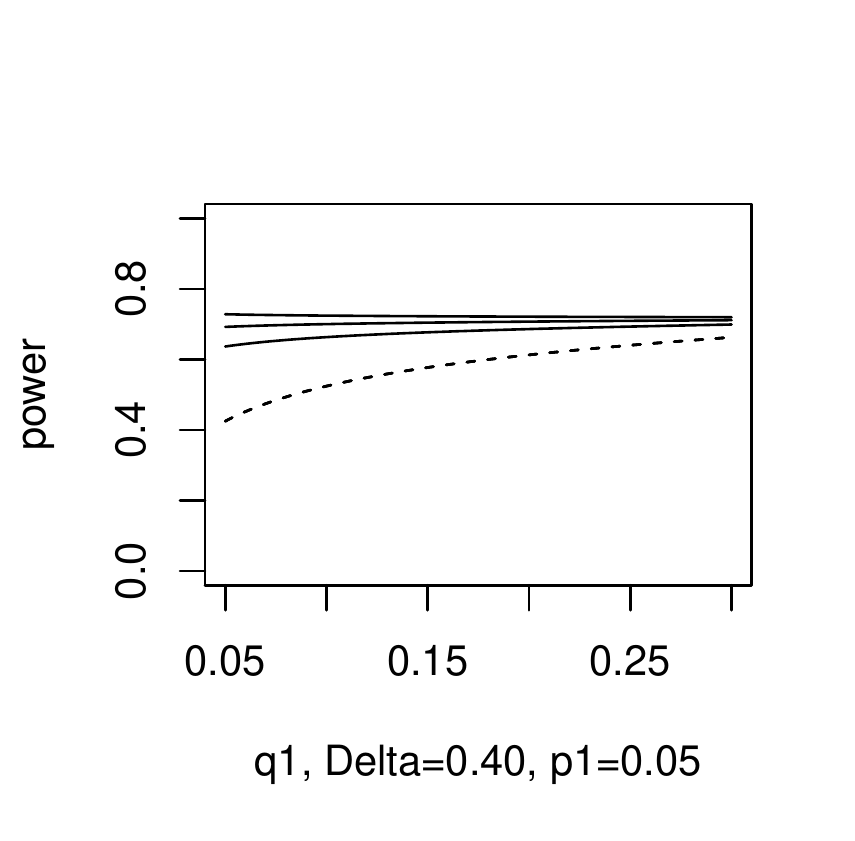}
\label{PowerpiCase}
\caption{Power function based on statistic $W$ as a function of $q_1$ for different values of $\pi$. 
From lowest continuous line to most upper line: $\hat\pi = 0.20$, $\hat \pi = \pi$, and $\hat\pi=0.075$. 
The dashed lines gives the power for the test based on test-statistic $T$.}
\end{figure}

\newpage

\noindent
{\underline{Power of $W_\delta$}}\\
In the previous paragraph we considered the effect of small deviations of $\delta$ near $\pi$ 
on the power of $W_\delta$. In this paragraph, we consider what happens if $\delta$ runs from 0 to 1. 
In the Figure 4 the asymptotic power of $W_\delta$ is plotted as a function of $q_1$, for different values of $\delta$.
The fat line (third line from above) indicates the power function for $W=W_\pi$ with $\pi=0.125$, 
and the thin continuous lines for different values
of $\delta$ ($\delta=0, 0.1, \ldots, 0.9, 1.0$, from upper to lowest line).
It can be seen that only for $\delta \approx \pi$ the lines are more or less constant; the power is not affected by the 
marker allele frequency. Above the fat line, the power functions are slightly decreasing as a function of $q_1$,
and below the fat line, the power is increasing. 
\begin{figure}[!p]
\centering  
\includegraphics[height=3in,width=4.5in]{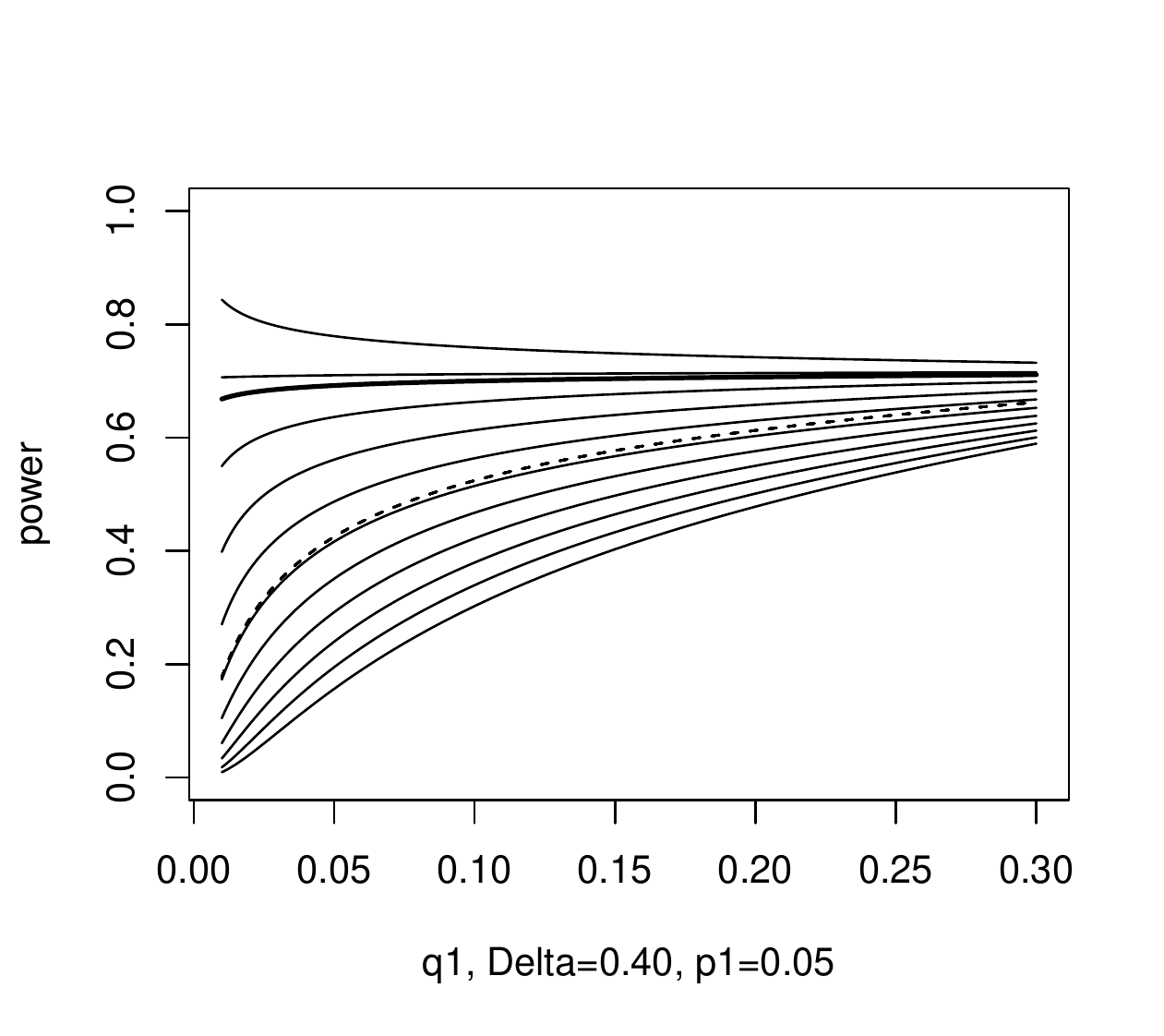}
\label{DeltaPlot}
\caption{Power function based on statistic $W_\delta$ as a function of $q_1$ for different values of $\delta$. 
Dashed lines: power based on $T$. Fat line (third lines from above): power based on $W=W_\pi$. The other lines are the power functions
based on $W_\delta$ for $\delta=0, 0.1, \ldots, 0.9, 1.0$, from upper to lowest line.
}
\end{figure}

\subsection{Finite samples}
The power function and the computation of p-values are based on asymptotic normality of the test-statistic $W$.
In practice the sample sizes are finite and not infinite. Therefore the normality approximation may not be exact and
a continuity correction may improve this approximation. A correction can be done in multiple ways. We used the following:
\begin{eqnarray}
W_{cor} = \sqrt{m} \; \frac{\hat q_{1|0} - \hat q_{1|1} \pm  \frac{1}{2} \min\{S, R \}/(2SR)}{\sqrt{\hat q_1 \hat q_2}},
\label{Wcor}
\end{eqnarray}
where the ``maximum'' could be replaced by the ``minimum'' if one prefers to correct less.
The null distribution of the test-statistic is asymptotically standard normal. In practice the number of observations 
is not infinite and the finite sample distribution may be different from the asymptotic distribution. 
We performed a simulation study to check the type I error for the three tests.
Like before we take $p_1=0.10$, $\pi_{11} = 0.60$ and $\pi_{22} = 0.10$ in the additive model.
Since we do the simulation under the null hypothesis $\Delta=0$.  
We simulate 100,000,000 times $R$ cases and $S$ controls
from the model given in Table 1, compute the test-statistic $T, W$ and $W_{cor}$, the corresponding p-values, and compute an estimate of
the type I error as the fraction of p-values smaller than $\alpha$. The results are given in Table 1.

\begin{table}[!h]
\centering
\begin{tabular}{|c|c|ccc|ccc|}
\hline
\multicolumn{2}{|c|}{} & \multicolumn{3}{c|}{$q_1=0.10$}  & \multicolumn{3}{c|}{$q_1=0.25$}    \\
\hline
\multicolumn{2}{|r|}{$\alpha$} & $1\times 10^{-3}$ & $1\times 10^{-4}$ & $1\times 10^{-5}$ & $1\times 10^{-3}$ & $1\times 10^{-4}$ & $1\times 10^{-5}$  \\
\hline
    & test & $\times 10^{-3}$ & $\times 10^{-4}$ & $\times 10^{-5}$  & $\times 10^{-3}$ & $\times 10^{-4}$ & $\times 10^{-5}$  \\ 
\hline    
$R=S=500$ &  &  &  &   &  &  &   \\ 
\hline
       & $T$          & 1.00124 & 0.9698  & 1.006 & 1.02640 & 1.0378 & 1.118      \\
$\delta=\pi$ & $W_{cor}$    & 1.12314  & 1.5258  & 2.324 &   0.97097 & 1.0419 & 1.175    \\
       & $W$          & 1.27542 & 1.7627 & 2.786 & 1.06058 & 1.1574 & 1.300    \\
$\delta=0.20$ & $W_{cor,\delta}$ & 1.04681 & 1.3322 & 1.893  & 0.95332 & 0.9981 & 1.111 \\
       & $W_{\delta}$ & 1.19209  & 1.5282  &  2.145  & 1.04465 & 1.1069 & 1.229  \\
$\delta=0.30$ & $W_{cor,\delta}$ & 0.93276  & 1.0104   & 1.226  & 0.92657 & 0.9375 & 0.969  \\
       & $W_{\delta}$ & 1.06701  & 1.1830  & 1.451   & 1.01572 & 1.0387 & 1.100   \\
$\delta=0.40$ & $W_{cor,\delta}$ & 0.87548  & 0.8479   & 0.899  & 0.91026 & 0.8918 & 0.922 \\
       & $W_{\delta}$ & 1.00305  & 0.9801  & 1.040   & 0.99397 & 0.9995 & 1.027 \\
\hline
$R=S=1000$ &  &  &  &   &  &  &   \\ 
\hline
       & $T$          &  1.00363 & 0.9930 & 1.023  & 1.02000 & 1.0359 & 1.005  \\
$\delta=\pi$ & $W_{cor}$    & 1.03673 & 1.2424 & 1.659 & 0.96959 & 1.0172 & 1.059 \\
       & $W$          & 1.12922 & 1.3821 & 1.845 & 1.03510 & 1.0996 & 1.155 \\
$\delta=0.20$ & $W_{cor,\delta}$ & 1.00093 & 1.1403 & 1.433  & 0.96411 & 0.9991 & 1.010  \\
       & $W_{\delta}$ & 1.09388 & 1.2723 & 1.607 & 1.02742 & 1.0754 & 1.104 \\
$\delta=0.30$ & $W_{cor,\delta}$ & 0.94062 & 0.9855 & 1.095 & 0.94886 & 0.9643 & 0.946  \\
       & $W_{\delta}$ & 1.03363 & 1.1026 & 1.228 & 1.01472 & 1.0400 & 1.025 \\
$\delta=0.40$ & $W_{cor,\delta}$ & 0.90898 & 0.8968 & 0.923 & 0.94026 & 0.9457 & 0.894 \\
       & $W_{\delta}$ & 0.99986 & 0.9976 & 1.041 & 1.00667 & 1.0146 & 0.976 \\
\hline
$R=S=2000$ &  &  &  &   &  &  &   \\ 
\hline
       & $T$          & 1.00585 & 1.0156 & 1.034 & 1.00851 & 1.0144  & 1.032 \\
$\delta=\pi$ & $W_{cor}$    & 1.00007 & 1.1170 & 1.368 & 0.96987  & 0.9840  & 1.025 \\
       & $W$          & 1.06733 & 1.2046 & 1.475 & 1.01565  & 1.0422  & 1.099  \\
$\delta=0.20$ & $W_{cor,\delta}$ & 0.98525 & 1.0636 & 1.245 & 0.96734  & 0.9793  & 0.988 \\
       & $W_{\delta}$ & 1.04942 & 1.1459 & 1.345  & 1.01188  & 1.0290  & 1.064 \\
$\delta=0.30$ & $W_{cor,\delta}$ & 0.95502 & 0.9889 & 1.075 & 0.96125  & 0.9620  & 0.990 \\
       & $W_{\delta}$ & 1.01906 & 1.0649 & 1.157 & 1.00730  & 1.0171  & 1.032 \\
$\delta=0.40$ & $W_{cor,\delta}$ & 0.93863 & 0.9400 & 0.973 & 0.95732  & 0.9536  & 0.971 \\
       & $W_{\delta}$ & 1.00167 & 1.0156 & 1.052 & 1.00206  & 1.0008  & 1.016 \\
\hline
\end{tabular}
\caption{Type I error for the three tests for several values of $\delta$,
different genetic models, and different sample sizes. ($\pi=0.15$)}
\end{table}

From the table it can be seen that the type I error for the test based on $W$ is slightly inflated especially if the 
sample sizes are small and the maf of the marker is low. This inflation disappears if the sample size or the maf grows.
After continuity correction as described in (\ref{Wcor}) there seems to be still an inflation, but this is smaller already.  
  
In the previous example the value of $\pi$ is 0.15. If the prevalence of the disease in the population is lower, the sample size
is small (around 500 cases and 500 controls) and the MAF of the marker is also quite low (0.10 or lower), the type I error inflation may become 
unacceptable. In that case one could decide to insert a higher value of $\delta$, what diminish the type I error, but decreases
the power of test under the alternative hypothesis.

We also performed several simulation studies under the alternative hypothesis. In all cases the power functions based on
finite samples were very similar to the asymptotic power function (results not shown).

\bigskip

\noindent
{\underline{Combined Test}}\\
We perform several simulation studies to study the performance of the combined test with test-statistic $U$.
The results are as expected. For large sample sizes the null distribution is close to a standard normal 
distribution (concluded from QQ-plots and histograms of the p-values) 
and the power-function equals the maximum of the power-functions for $T$ and $W$.
For low sample sizes and/or low minor allele frequencies at the marker, the null distribution deviates 
from the standard normal distribution in just one tail.
For all observations in that tail $\hat Q<1$ (the test-statistic $U$ equals $W$). This was also seen for the 
test-statistic $W$ and is therefore as expected. This problem can be easily solved by recalculating the small
p-values in the tail by permutations.

The observations in the opposite tail all had $\hat Q >1$ ($U$ equals $T$) and the fit with the standard normal 
distribution is good (as expected, since the null distribution
for $T$ is standard normal also if the sample size is low).
The results of these simulation studies are not shown in this paper, because, to our opinion, the results are 
as expected and do not add much to the paper.

\section{Discussion}
Several methods have been proposed for selecting markers for follow-up in a case-control association study.
One of the most popular test is the allele based test that considers the difference of marker allele frequency 
among cases and controls. A disadvantage of this test is its preference for markers with high minor allele frequencies;
markers with low minor allele frequency have less power to be detected than markers with a high minor allele frequency.

In this paper a new allele based association test for finding markers that are associated with a disease
is proposed. The test is model-free and the test-statistic can be computed easily and fast. Moreover, 
the asymptotic null distribution of the test-statistic is known what makes the computations of p-values fast; 
permutations are not necessary. 
The power of the test is higher than for the commonly used test for many genetic models that are of practical interest. 
For those models, the lower the minor allele frequency is, the more power is gained by using the new test.
This is mainly caused by the fact that the power of the proposed test is approximately constant as a function of the 
marker allele frequency; the proposed test does not favor markers with a high minor allele frequencies. 
So, ranking the markers based on their p-values becomes a more objective way of selecting
interesting markers for follow-up studies and, because of its high power for markers with a low minor allele 
frequency (compared to the existing test), the proposed test may reveal interesting regions for follow-up study.   
 
The proposed test-statistic depends on the parameter $\pi$, the prevalence of the disease. 
This parameter cannot be estimated based on the data itself, but should be 
estimated from external data. For most diseases, population risks are available, for instance from national
registries. In the paper it is shown that the type I error is hardly affected by misspecification of $\pi$, because
under the null hypothesis the parameter (almost) drops out from the test-statistic if the number of observations is large. 
In practice, it is quite common to estimate nuisance parameters in the model based on external data and to assume that
these estimated nuisance parameters are known when performing statistical tests or constructing confidence intervals 
for the parameter of interest. The extra uncertainty due to the estimation of these nuisance parameter are often not taken into
account what may lead to wrong type I errors, in general . In \cite{Jonker} the effect on the type I error for 
the likelihood ratio test-statistic is studied. 

We generalized the model by allowing other values for $\pi$ between zero and one. This may yield higher power for some
markers, but the a priori preference for markers with high or low MAFs is back again. Moreover, by taking
values near the boundary (zero or one), the type I error may increase above acceptable levels.
We therefore advice to use $\pi$ as first choice. However, if $\pi$ itself is low (near zero) and the sample size is not 
huge, the type I error may be inflated if the marker MAF is low. In that case one could decide to impute a higher 
value for $\delta$ in the denominator of the test-statistic, so $\delta>\pi$, to give up some power and lower the number 
of false discoveries. This was seen in the simulation study in this manuscript.

In the derivation of the test-statistic $W$ as well as in the simulation studies we assume that the alleles 
at a marker and causal variant are in Hardy-Weinberg equilibrium. However, deviations from Hardy-Weinberg equilibrium 
can inflate the chance of a false-positive association (\cite{Schaid}).
In \cite{Schaid} an test-statistic that accounts for deviation from HWE is introduced;
an extra term is added to the variance in the denominator of the test-statistic. Our test-statistic
can be adjusted in a similar way, what makes the type I error of the test robust against deviation from Hardy Weinberg.

When the Hardy-Weinberg proportions hold in the total population, the allele-based test $T$ and
the Cochran-Armitage Trend test for the additive model are asymptotically equivalent under the null hypothesis 
(\cite{Armitage1955,BookAsso}). 
Since $T$ and $W$ are also asymptotically equivalent under the null hypothesis, this also
holds for $W$ and the Cochran-Armitage Trend test (CATT). 
Under the alternative hypothesis, the power functions
differ. Based on a simulation study, \cite{BookAsso} show that the power 
of the allele-based test $T$ and the CATT for additive models are, nevertheless, very similar, 
with a slightly higher power
for the allele-based test $T$ under the recessive model and for the CATT under the dominant model
(for genetic models they consider). For the genetic models for which the 
test based on $W$ is more powerful than the test based on $T$, the test based on $W$ is also more 
powerful than the CATT.      

Another association test is a score test based on a logistic regression model. 
The score test statistic equals the CATT (\cite{BookAsso}), from which it directly follows 
from the previous paragraph that in many interesting settings, the allele-based test $W$ is also 
more powerful than the score-test for a logistic regression model. 

For some genetic settings the test based on $T$ is more powerful than the test based on $W$.
We therefore combined the two test-statistics to a test-statistic $U$ that always has an (asymptotic) power
of at least the tests $T$ or $U$.  Although the test-statistic is more complicated now, it still has
asymptotically a standard normal distribution under the null hypothesis and p-values can be easily obtained. 

A part of the missing heritability might be explained by causal variants with a low minor allele frequency. 
The test proposed in this paper, 
may help detecting a part of the undiscovered causal variants. 


\bigskip

\addcontentsline{toc}{section}{Bibliography}
\bibliographystyle{plain}
\bibliography{ref}

\begin{thebibliography}{10}

\bibitem{Armitage1955}
P~Armitage.
\newblock Tests for linear trends in proportions and frequencies.
\newblock {\em Biometrics}, 11:375--386, 1955.

\bibitem{Balding2006}
DJ~Balding.
\newblock A tutorial on statistical methods for population association studies.
\newblock {\em Nat Rev Genet}, 7:781--791, 2006.

\bibitem{Devlin}
B~Devlin and N~Risch.
\newblock A comparison of linkage disequilibrium measures for fine-scale
  mapping.
\newblock {\em Genomics}, 29:311--322, 1995.

\bibitem{Jonker}
MA~Jonker and AW~van~der Vaart.
\newblock On the correction of the asymptotic distribution of the likelihood
  ratio statistic if nuisance parameters are estimated based on an external
  source.
\newblock {\em Int J of Biostat}, 10(2):123--142, 2014.

\bibitem{Manolio}
TA~Manolio, FS~Collins, NJ~Cox, DB~Goldstein, LA~Hindorff, DJ~Hunter,
  MI~McCarthy, EM~Ramos, LR~Cardon, A~Chakravarti, JH~Cho, AE~Guttmacher,
  A~Kong, L~Kruglyak, E~Mardis, CN~Rotimi, M~Slatkin, D~Valle, AS~Whittemore,
  M~Boehnke, AG~Clark, EE~Eichler, G~Gibson, JL~Haines, TFC Mackay,
  SA~McCarroll, and PM~Visscher.
\newblock Finding the missing heritability of complex diseases.
\newblock {\em Nature}, 461:747--753, 2009.

\bibitem{Schaid}
DJ~Schaid and SJ~Jacobsen.
\newblock Biased tests of association: of allele frequencies when departing
  from hardy-weinberg proportions.
\newblock {\em American Journal of Epidemiology}, 149(8):706--711, 1999.

\bibitem{Schork}
NJ~Schork, SS~Murray, KA~Frazer, and EJ~Topol.
\newblock Common vs. rare allele hypotheses for complex diseases.
\newblock {\em Current Opinion in Genetics and Development}, 19:212--219, 2009.

\bibitem{Stromberg2008}
U~Stromberg, J~Bjork, K~Broberg, F~Mertens, and P~Vineis.
\newblock Selection of influential genetic markers among a large number of
  candidatesbased on effect estimation rather than hypothesis testing: an
  approach for genome-wide association studies.
\newblock {\em Epidemiology}, 19:302--8, 2008.

\bibitem{Stromberg2009}
U~Stromberg, J~Bjork, P~Vineis, K~Broberg, and E~Zeggini.
\newblock Ranking of genome-wide association scan signals by different
  measures.
\newblock {\em Int J Epidemiology}, 38:1364--1373, 2009.

\bibitem{vanderVaart}
AW~van~der Vaart.
\newblock {\em Asymptotic Statistics}.
\newblock Cambridge University Press, Cambridge, 1998.

\bibitem{W2004}
S~Wacholder, S~Chanock, M~Garcia-Closas, L~El-ghormli, and N~Rothman.
\newblock Assessing the probability that a positive report is false: an
  approach for molecular epidemiology studies.
\newblock {\em J Nat Cancer Inst}, 96:434--41, 2004.

\bibitem{Wakefield2008}
J~Wakefield.
\newblock Reporting and interpretation in genomewide association studies.
\newblock {\em Int J Epidemiology}, 37:641--53, 2008.

\bibitem{Wakefield2009}
JA~Wakefield.
\newblock Bayes factors for genome-wide association studies: comparison with
  p-values.
\newblock {\em Genet Epidemiol}, 33:79--86, 2009.

\bibitem{Wellek2012}
S~Wellek and A~Ziegler.
\newblock Cochran-armitage test versus logistic regression in the analysis of
  genetic association studies.
\newblock {\em Hum her}, 73:14--17, 2012.

\bibitem{BookAsso}
G~Zheng, Y~Yang, X~Zhu, and R~Elston.
\newblock {\em Analysis of Genetic Association Studies}.
\newblock Springer, 2012.

\end{thebibliography}

\section{Appendix}
\subsection{Appendix A: derivation of the equality (\ref{pMLUDelta})}
\subsection{Appendix B: The fraction $Q$}

\end{document}